# Surface Grafting of Graphene Flakes with Fluorescent Dyes: A Tailored Functionalization Approach

Ylea Vlamidis [1,2,*], Carmela Marinelli [2], Aldo Moscardini [3], Paolo Faraci [3], Stefan Heun [1,*] and Stefano Veronesi [1]

[1] NEST, Istituto Nanoscienze–CNR and Scuola Normale Superiore, Piazza San Silvestro 12, 56127 Pisa, Italy; stefano.veronesi@nano.cnr.it
[2] Department of Physical Science, Earth, and Environment, University of Siena, via Roma 56, 53100 Siena, Italy; carmela.marinelli@unisi.it
[3] Scuola Normale Superiore, Laboratorio NEST, Piazza San Silvestro 12, 56127 Pisa, Italy; aldo.moscardini@sns.it; paolo.faraci@sns.it
* Correspondence: ylea.vlamidis@unisi.it; stefan.heun@nano.cnr.it

**Abstract:** The controlled functionalization of graphene is critical for tuning and enhancing its properties, thereby expanding its potential applications. Covalent functionalization offers a deeper tuning of the geometric and electronic structure of graphene compared to non-covalent methods; however, the existing techniques involve side reactions and spatially uncontrolled functionalization, pushing research toward more selective and controlled methods. A promising approach is 1,3-dipolar cycloaddition, successfully utilized with carbon nanotubes. In the present work, this method has been extended to graphene flakes with low defect concentration. A key innovation is the use of a custom-synthesized ylide with a protected amine group (Boc), facilitating subsequent attachment of functional molecules. Indeed, after Boc cleavage, fluorescent dyes (Atto 425, 465, and 633) were covalently linked via NHS ester derivatization. This approach represents a highly selective method of minimizing structural damage. Successful functionalization was demonstrated by Raman spectroscopy, photoluminescence spectroscopy, and confocal microscopy, confirming the effectiveness of the method. This novel approach offers a versatile platform, enabling its use in biological imaging, sensing, and advanced nanodevices. The method paves the way for the development of sensors and devices capable of anchoring a wide range of molecules, including quantum dots and nanoparticles. Therefore, it represents a significant advancement in graphene-based technologies.

## 1. Introduction

Graphene, a single-atom-thick planar sheet of sp$^2$-bonded carbon atoms arranged in a honeycomb crystal lattice, has garnered significant attention over the last two decades after its first isolation occurring as recently as 2004 [1]. This groundbreaking discovery has opened up new opportunities for research in physics, chemistry, biotechnology, and materials science. Graphene boasts remarkable properties including exceptional electrical conductivity, mechanical flexibility, optical transparency, thermal conductivity, and low coefficients of thermal expansion [2,3]. These unique characteristics have sparked considerable interest across academia and industry, spanning various fields in the areas of polymer nanocomposites [4], supercapacitor devices [5], drug delivery systems [6], solar cells [7], memory devices [8], field-effect transistor devices [9], biosensors [10], and more. Moreover, the possibility to precisely control the functionalization or chemical derivatization of graphene and graphene oxide (GO) opens new avenues for tailoring and enhancing their properties, thereby expanding their potential applications across various fields [11,12]. Recent advances in the robust and well-controlled functionalization of graphene and graphene oxide have paved the way for a wide range of applications, including aptamer-based biosensors [13], gas sensing [14], and next-generation electrochemical, electronic, and optoelectronic devices [15]. These breakthroughs highlight the immense potential of functionalized graphene in both fundamental research and industrial applications.

In particular, the covalent functionalization of graphene offers several advantages as it enables significantly stronger modifications of the geometric and electronic structure of graphene [16,17]. This method allows to tune the properties of graphene, like enhancing its stability in hydrophilic/hydrophobic media [18,19], combining graphene properties with those of other compound classes possessing specific functions, enabling the modulation of graphene's molecular-level doping, or systematically tuning the Fermi level by the introduction of electron-donating/electron-withdrawing groups (n-doping or p-doping) [20]. For these reasons, achieving an effective method for producing surface-functionalized graphene sheets on a large scale has become a major objective for many researchers.

The covalent functionalization of graphene, which induces a hybridization change in its carbon atoms from $sp^2$ to $sp^3$, can be effectively achieved through several reaction mechanisms, with radical and cycloaddition reactions being among the most efficient approaches.

Radical reactions occur through the use of aryl diazonium salts to generate radicals upon electrochemical reduction or through thermal or photoinduced decomposition [21,22]. This functionalization process is triggered by a single-electron transfer from the graphene lattice to the aryl diazonium salt, leading to the formation of an aryl radical via the release of a $N_2$ molecule. Cycloaddition reactions provide an alternative and highly effective method of functionalizing graphene, involving the simultaneous formation of two σ bonds with $sp^3$ carbons in the graphene lattice. A unique feature of these reactions is that, due to the degeneracy of electronic states at the Dirac point, graphene can act as various synthons (e.g., diene, allyl, etc.), allowing for versatile reactivity in cycloaddition processes [23]. Cycloaddition can be achieved using reactive intermediates such as nitrenes, carbenes, and arynes [24–26]. These species covalently modify graphene through CH insertion or cycloaddition reactions. However, this type of reaction, which employs highly reactive intermediates, presents several disadvantages, including the occurrence of undesired side reactions that can disrupt the graphene lattice and the challenge of achieving controlled selectivity, which often leads to spatially uncontrolled functionalization. Additionally, the short lifespan of radical or intermediate species poses a significant challenge as they can decompose rapidly, necessitating precise control over reaction conditions, such as temperature and light, to ensure effective and consistent functionalization.

Currently, one of the most promising routes for the covalent functionalization of graphene is the 1,3-dipolar cycloaddition (1,3-DC) of azomethine ylide, as it is a more selective and controlled method [27]. This approach has been already explored for the chemical modification of carbon nanotubes, fullerenes, and other carbon nanostructures [28–30]. The versatility of this method is attributed to the wide range of organic derivatives available through the selection of appropriate precursors. A recent comprehensive investigation has been conducted on the grafting of azomethine ylide onto both graphene nanosheets and reduced graphene oxide via 1,3-dipolar cycloaddition, which included Density Functional Theory (DFT) simulations and investigation of the influence of organic solvents on dispersion properties [31,32].

Here, we demonstrate the possibility to exploit this reaction mechanism to successfully functionalize graphene flakes with specific molecules, such as well-known commercial fluorescence markers for biological and environmental applications [33–35]. Small molecule organic fluorescent dyes are primary reagents in scientific research, particularly employed in cells, tissues, microorganisms, and the environment for various applications such as imaging, sensing, and drug delivery [36]. The advantage offered by the incorporation of fluorescent dyes in graphene nanosheets is the direct and fast detection of the functionalized samples by excitation of fluorescence, besides the possibility to localize the most reactive sites (defects) through fluorescence imaging. The literature reports the functionalization of graphene oxide through covalent bonding with fluorescent dyes, exploiting the reactivity of oxygen-containing groups [37]. It is well known that the presence of functional groups, as well as lattice defects such as $sp^3$ carbon hybridization, vacancy-like defects, and boundary-like defects, can significantly enhance the reactivity of the carbon atoms in graphene, thus favoring the covalent functionalization with organic molecules [16,38]. Nevertheless, defects in the graphene crystal lattice lead to an undesired degradation of its properties such as electrical conductivity, mechanical strength, and optical characteristics. On the contrary, achieving the functionalization of graphene with low defect density, essential for any application, is still challenging. For this reason, we investigated the covalent functionalization of graphene nanosheets, which, due to their nature, feature low defect density, mainly at the edges.

This work introduces a novel method for the covalent functionalization of graphene flakes with a low defect concentration, marking a significant improvement that enables the attachment of specific molecules such as fluorescence markers commonly used in biological research. A central innovation is the use of a custom-synthesized ylide with a protected amine group tert-butoxycarbonyl (Boc), allowing for further functionalization with target molecules. In contrast to conventional methods that often damage the structure of graphene, this approach minimizes defects while enabling effective functionalization thanks to the high selectivity of the method employed. The first step of functionalization, involving the grafting of the ylide on the graphene flakes via 1,3-DC, was confirmed by Raman spectroscopy. Next, after Boc cleavage, the fluorophores were linked to the ylide. The fluorophores chosen contain a N-hydroxysuccinimide (NHS) ester terminal group, which is frequently employed for the covalent conjugation of amine-containing biomolecules (e.g., proteins, peptides, or to label a protein or peptide) to surfaces through amide linkage [39]. The active ester of the employed fluorophores (Atto 425, Atto 465, and Atto 633 NHS ester) can react with primary amines in slightly alkaline conditions (pH 7.2 to 9) to yield stable amide bonds.

Photoluminescence spectroscopy and confocal microscopy were utilized to detect graphene flakes labeled with fluorescent dyes, enabling the precise characterization of the samples and demonstrating the successful ylide 1,3-dipolar cycloaddition onto graphene, followed by the attachment of fluorescent dyes through NHS ester derivatization.

This approach paves the way for sensor development and device fabrication, providing a flexible platform for anchoring not only fluorescent dyes but also quantum dots, nanoparticles, and other molecules. With broad potential for applications in biological imaging, sensing, and advanced nanodevices, this technique represents a significant advancement in the functionalization of graphene-based materials. In these applications, toxicity is a key point.

While some studies suggest a promising biocompatibility of graphene, the toxicity of functionalized graphene remains highly dependent on surface chemistry, functionalization strategy, and potential contaminants [40]. Therefore, further comprehensive studies are essential to ensure its safe and effective use in biomedical applications.

## 2. Materials and Methods

### 2.1. Chemicals

BeDimensional S.p.A. (Italy) supplied the wet-jet milled exfoliated graphene powder (99%). Ethanol (≥96%), phosphoric acid (85%), N,N-dimethylformamide (DFM, anhydrous, 99.8%), N-methylglycine (98%), di-tert-butyl dicarbonate ($Boc_2O$, 99%), N,N′-Diisopropylcarbodiimide (DIC, 99%), triethylamine ($Et_3N$, >99.5%), hexamethylenediamine (98%), acetonitrile anhydrous (MeCN, 99.8%), dichloromethane (99.8%), methanol (99.8%), anhydrous dimethyl sulfoxide (DMSO, ≥99.9%), 3,4-dihydroxybenzaldehyde (97%), and benzyl-2-bromoacetate (98%) were purchased from Sigma-Aldrich (Darmstadt, Germany). Sodium carbonate (for analysis, ≥99.5%) was purchased from Carlo Erba (Milano, Italy), while Atto 425 NHS ester (>90%), Atto 465 NHS ester (>90%), Atto 633 NHS ester (>90%) were supplied by ATTO-TEC GmbH (Siegen, Germany).

### 2.2. Synthesis of the Linker Molecule

The proposed approach for the functionalization of graphene flakes was implemented through a multi-step process, with a detailed schematic flowchart outlining the key phases presented in Figure 1. This visual representation provides a clear overview of each stage, enhancing the understanding of the methodology.

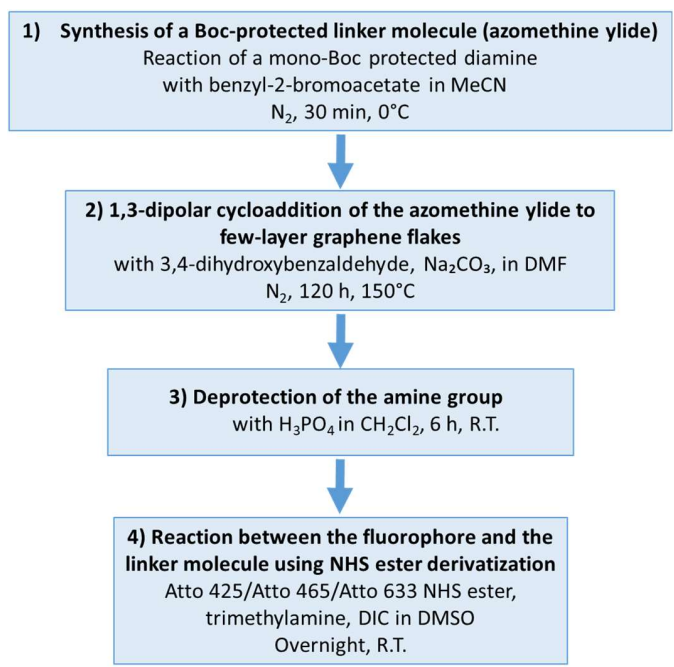

**Figure 1.** Schematic flowchart describing the multi-step functionalization of graphene flakes with fluorescent molecules.

In this work, the mono-functionalization of diamine is key in holding functional groups at one end while the other end is free to attach to the graphene flakes. Therefore, we synthesized a mono-protected diamine starting from 1,6-hexanediamine and using tert-butyloxycarbonyl (Boc) as an amine protecting group [41].

First, the two amine groups of hexamethylenediamine are differentiated as an acid salt and a free base, which is ready for further functionalization (Scheme 1) [42]. Selective protonation of one amine at moderately acidic pH occurs due to inductive and electrostatic effects, ensuring that only one amine forms a salt while the other remains a reactive free base. Protonation of the first amine withdraws electron density along the aliphatic chain, lowering the basicity of the second amine and reducing its likelihood of protonation. Additionally, electrostatic repulsion between two $NH_3^+$ groups further disfavors double protonation under these conditions. Gaseous hydrochloric acid was obtained by a reaction between HCl (0.05 mol) and $H_2SO_4$ (0.05 mol) at 40 °C under a nitrogen atmosphere. HCl was directly bubbled into a flask containing 40 mL of methanol. Hexamethylenediamine (0.05 mol) was dissolved in 20 mL of methanol, and, after 30 min, was slowly added to the reaction flask through a dropping funnel over a 20 min period. The solution was placed in a cooling bath (acetone/ice/NaCl) and stirred overnight. The mono-salified diamine produced from this reaction was dried using a rotary evaporator and subsequently purified by HPLC-MS to eliminate any residual free diamine, ensuring a high-purity final product.

The mono-salified diamine (1.8 mmol) was then dissolved in 40 mL of methanol, and a solution containing $Boc_2O$ (1 eq.) in 40 mL of methanol was slowly added to let the free amine react, leading to the formation of the mono-Boc protected diamine. Once the addition was complete, the solution was stirred for about 1 h in a cooling bath under a nitrogen atmosphere. This method has demonstrated high efficiency in contrast to the conventional approach which requires careful control over reagents concentration and a gradual addition of $Boc_2O$ solution over a long period of time to prevent di-Boc formation [43,44]. Once the reaction was completed, the solvent was evaporated, and diethyl ether was added to remove the unreacted diamine residue. Then, 1M NaOH solution was added to obtain the amine salt free. The reaction product was then extracted by organic solvent, washed in dichloromethane, and dried using rotary evaporator equipment.

The mono-Boc diamine (6 mmol) was then dissolved in 40 mL of anhydrous acetonitrile in a three-neck round-bottomed flask equipped with a magnetic stirring bar. Then, a solution of benzyl-2-bromoacetate (alpha-bromo derivative, 2 mmol) and trimethylamine (4 mmol) in anhydrous acetonitrile (20 mL) was slowly added to the reaction flask through a dropping funnel. The reaction was carried out under nitrogen atmosphere, and after 30 min, the reaction was complete. Finally, the product was analyzed and purified using a tandem UHPLC-MS system (refer to Figure S1 in the Supporting Information) before it was freeze-dried and stored at −20 °C.

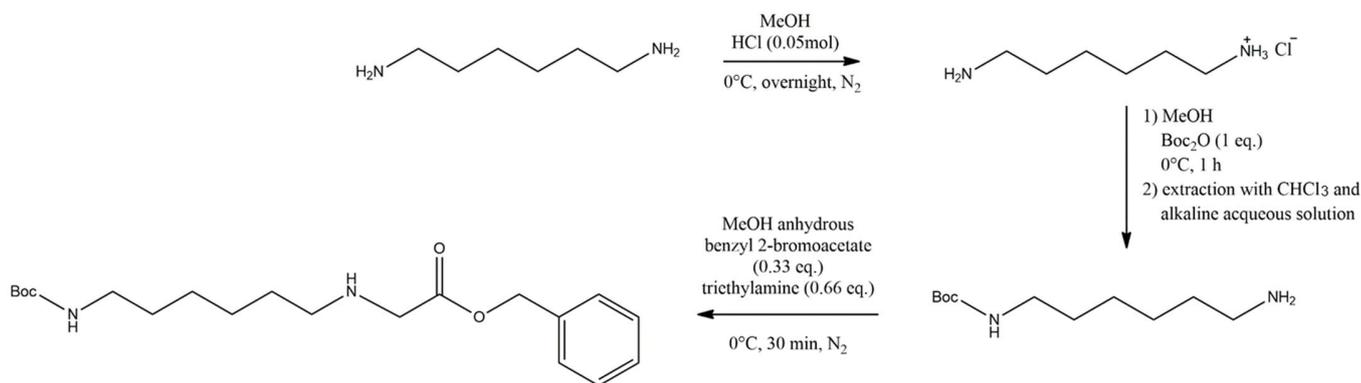

**Scheme 1.** Schematic of synthesis of the linker molecule (Boc-protected amino groups) from hexamethylenediamine.

*2.3. Covalent Functionalization of Graphene Flakes by 1,3-Dipolar Cycloaddition*

The first step involves the covalent functionalization between the organic linker and the graphene flakes, as described in Scheme 2. For this reaction, 3,4-dihydroxybenzaldehyde (6 mg, 0.18 mmol), sodium carbonate (5 mg), and the previously synthesized linker molecule (6 mg, 0.28 mmol) were added to the few-layers graphene suspension (2 mg) in DMF. Specifically, sodium carbonate (which reduces the acidity of the reaction environment) and 3,4-dihydroxybenzaldehyde are necessary for the in-situ formation of the 1,3-dipolar compound (azomethine ylide), which ultimately binds to graphene through the formation of a pyrrolidine ring (refer to Scheme 2). In the reaction mechanism, the formation of the imine (reaction intermediate) increases the acidity of the alpha proton, leading to the formation of a carbanion (negatively charged carbon atom). The shift in the double bond, in turn, results in the formation of the 1,3-dipoles that represent the reactive species.

The reaction was carried out for 120 h at 150 °C under magnetic stirring [31,32]. Fresh excess reagents were added daily to drive the reaction and increase the yields of graphene functionalization, as the ylide is susceptible to deactivation through side reactions with 3,4-dihydroxybenzaldehyde or even through self-reactivity with another ylide molecule. To avoid secondary reactions resulting from solvent oxidation at elevated temperatures, the reaction was conducted under a nitrogen atmosphere. The resulting suspension was then washed several times with clean solvent. The mixture was centrifuged at 13,000 rpm, the solvent was removed, clean solvent was added, and then the flakes were re-suspended by sonication. This procedure was repeated several times using DMF, water, acetone, and eventually chloroform.

The second step involves the deprotection of the amine group. Conventional methods for N-Boc deprotection primarily rely on cleavage in acidic conditions [45]. The Boc protecting group was cleaved, suspending the graphene flakes functionalized with the linker in dichloromethane (600 μL), and then phosphoric acid (15 eq.) was added. The mixture was stirred for about 6 h at room temperature and checked periodically by FT-IR spectroscopy to confirm the completion of the reaction (refer to Supporting Information Figure S2).

Eventually, the flakes were rinsed in dichloromethane and methanol by sequentially centrifugation and resuspension.

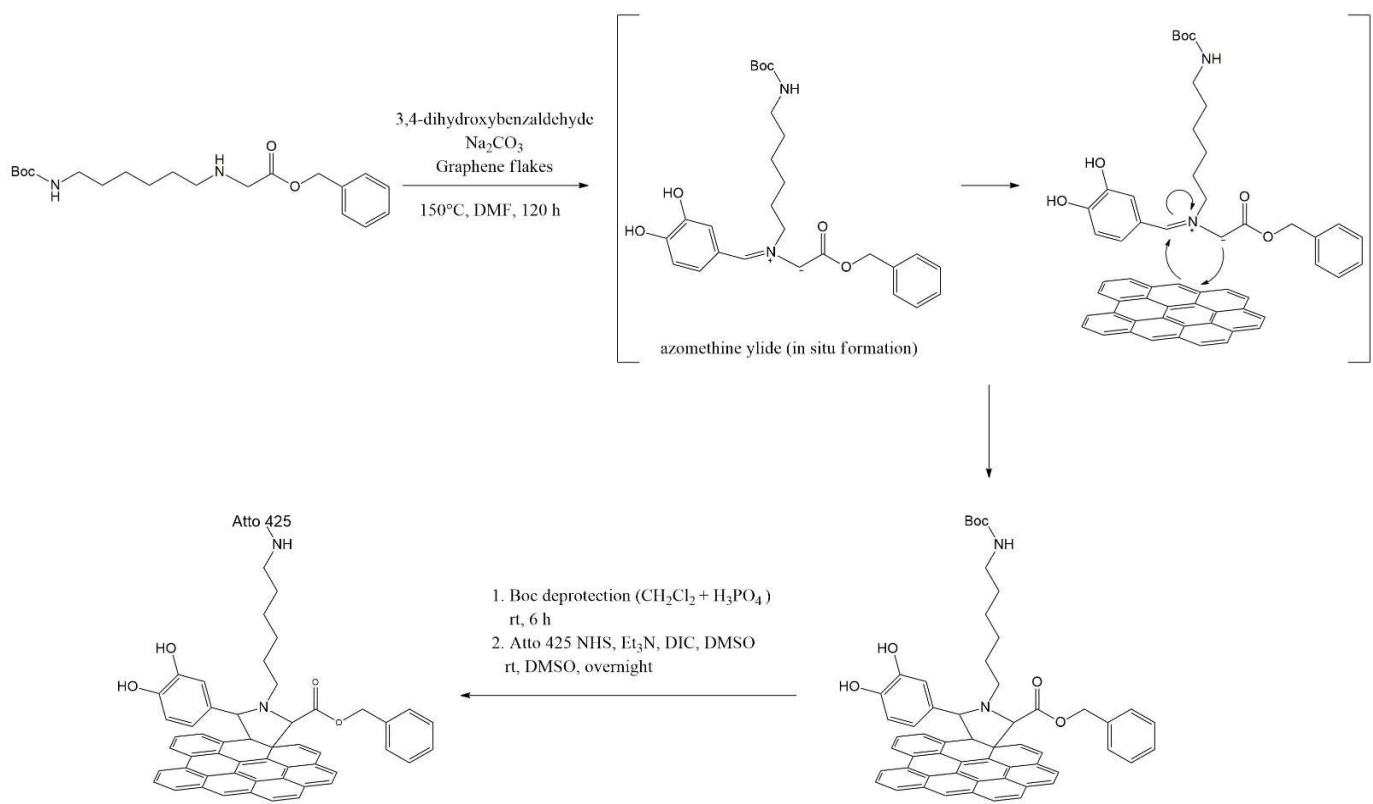

**Scheme 2.** Schematic representation of the functionalization reactions: reaction mechanism of the 1,3-DC between the linker and the graphene flakes and attachment of a fluorophore to the linker molecule after deprotection of the amino group.

Once the amine group was deprotected, the fluorophore was linked to the amine. This third reaction step was carried out suspending the graphene flakes-linker in 500 μL of DMSO containing the desired fluorophore (10 μg). In the reaction ambient, triethylamine was added to reach a mild alkaline pH, and DIC (100 μL) was used to promote amidation reaction by the activation of the carboxyl group [46]. The mixture was stirred overnight at room temperature, and the reaction flask was carefully protected from light. The graphene flakes were rinsed in DMSO, acetonitrile, and acetone to remove the excess of reagents and catalyst. The fluorophore-labeled graphene flakes were dried and stored at −20 °C in a dark vial bottle.

*2.4. Spectroscopic, Optical, and Morphological Characterization Techniques*

For the spectroscopic, optical, and morphological characterization, 5 μL of the graphene flakes suspended in chloroform were applied onto clean silicon substrates with native oxide (Si/SiO$_2$) by the drop-casting method and dried under nitrogen flux.

Raman spectroscopy was performed using a Renishaw InVia system featuring a confocal microscope, an excitation laser with a wavelength of 532 nm (2.62 eV), and an 1800 l/mm grating (spectral resolution of 2 cm$^{-1}$). All spectra were acquired with a 100× objective (NA = 0.85, spot size ~1 μm) using the following parameters: excitation laser power ~90 mJ μm$^{-2}$, two acquisitions per spectrum. A comprehensive statistical analysis was performed on 75

Raman spectra collected from the surfaces of different graphene flakes, both for pristine and functionalized graphene, providing robust insight into the uniformity and consistency of the functionalization.

Photoluminescence spectra were acquired with an excitation wavelength of 473 nm using the Raman spectrometer, with 2400 l/mm grating, 50× objective, and ~6 mJ μm$^{-2}$ laser power.

Fluorescence imaging and lifetime measurements were performed with a confocal microscope (Leica Microsystems, Wetzlar, Germany), equipped with a 63×, oil immersion objective (Leica Microsystems). Pulsed diode lasers operating at a frequency of 40 MHz were employed for excitation, and each sample was excited as close as possible to the wavelength of maximum absorption of the fluorophore molecules under investigation. Fluorescence lifetime imaging microscopy (FLIM) was employed to provide insights into the local microenvironment and molecular interactions, as fluorescence decay patterns are significantly influenced by factors such as local pH changes, energy transfer, and molecular binding [47].

Fluorescence microscopy images were analyzed using ImageJ software. For each image, the color of the pixels corresponds to the perceived color of the fluorescence emission (spectral emission wavelength).

The atomic force microscopy (AFM) data were acquired with a Bruker Dimension Icon system, operating in tapping mode, and Gwyddion software was used for processing the images.

Scanning electron microscopy (SEM) images were acquired with a Jeol JSM-7500F instrument using the following parameters: working distance of 6 mm, acceleration voltage of 5 kV, and emission current of 20 μA.

## 3. Results and Discussion

*3.1. Surface Characterization*

Graphene flakes produced by wet-jet milling [48] were dispersed in DMF by homogenization in order to obtain a stable dispersion (~0.2 mg mL$^{-1}$) [31]. Such graphene dispersion exhibits stacks of several overlapping layers with a smooth planar structure, showing some scrolling on the edges of the graphene (refer to AFM and SEM characterizations in Figure 2). According to the morphological characterization, the nanosheets display different lateral sizes, ranging from a few hundred nanometers to ~3 μm. In the AFM profile obtained from a flat area, heights corresponding to a few layers of graphene can be identified, with thickness ranges from about 0.6 to 1 nm (Figure 2b).

*3.2. Raman Characterization of Graphene Functionalized with the Linker by 1,3-Dipolar Cycloaddition*

The first step of functionalization, i.e., the grafting of the linker molecule on graphene flakes by cycloaddition, was confirmed by Raman analysis before and after the cycloaddition. Figure 3 shows representative Raman spectra of the pristine graphene flakes compared to those of the graphene-linker sample, along with a typical optical image. The typical features in the Raman spectrum of graphene are the G band and the 2D band, centered at 1581.6 cm$^{-1}$ and 2720.8 cm$^{-1}$, respectively. The 2D band shows a full width at a half maximum (FWHM) of about 64 cm$^{-1}$, which allows us to identify the graphene flakes as few- layer graphene [49]. The D peak, centered at 1350.5 cm$^{-1}$, and the D′ peak, here centered at 1616 cm$^{-1}$, indicate the presence of defects in the lattice, which breaks the symmetry of the carbon honeycomb lattice. For low defect concentrations, the D peak intensity is proportional to the amount of defects [50]. For pristine graphene, the ratio I(D)/I(D′) is ~4.8, suggesting the presence of boundary-like and vacancy-like defects in the sample [51].

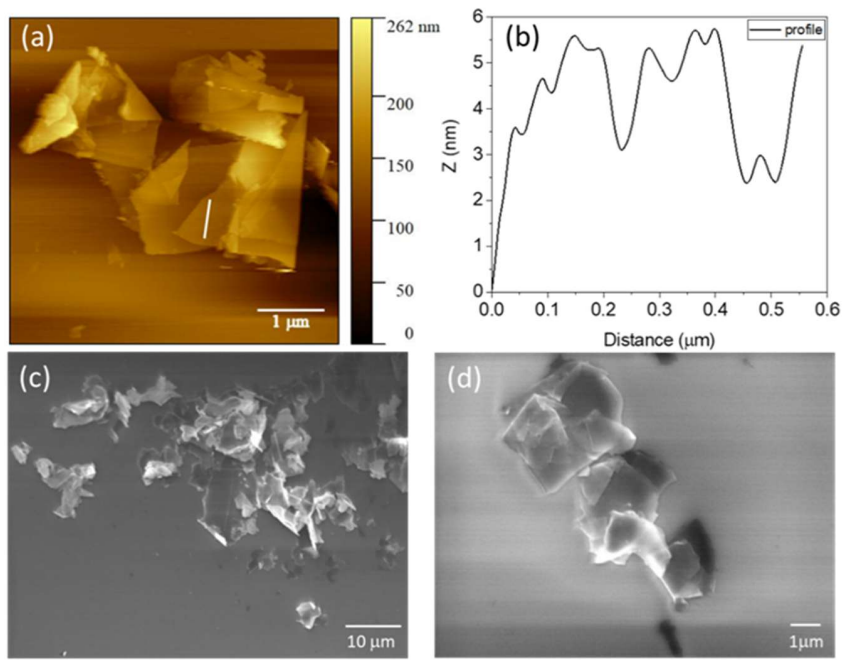

**Figure 2.** (**a**) AFM topography with (**b**) height profile along the white line indicated in (**a**). (**c**,**d**) SEM images at different magnification of the graphene flakes drop-cast on Si/SiO$_2$ from a chloroform suspension.

The Raman spectrum of functionalized graphene (Figure 3a) shows clearly a decreased intensity of the D peak compared to the pristine sample. The defect density can be assessed by the peak intensity ratio between the D band and the G band. The inset of Figure 3a presents the statistical analysis of 75 Raman spectra acquired on the surfaces of different few-layer graphene flakes of both pristine and functionalized samples, providing a detailed comparison of the defect density. Here, the ratio I(D)/I(G) varies from an initial average value of ~0.11 for the pristine sample to a value of ~0.022 for functionalized graphene flakes. This decrease in I(D)/I(G) can be explained considering that the linker molecule grafts onto graphene's most favorable bonding sites, which are the lattice defects, possibly leading to a local structural relaxation and a decrease in the Raman intensity of the defects. Considering that the lateral size of the graphene flakes is comparable with the Raman laser spot size (~1 µm), edge defects contribute significantly to the intensity of the D peak of the pristine graphene flakes, which, after functionalization, are saturated by the presence of the molecules, leading to a decrease in the intensity of the D peak [31,52]. On the contrary, the intensity and the position of the 2D band are not greatly affected compared to pristine graphene, confirming that the process does not significantly modify the structural order of the graphene lattice, and no disorder due to amorphous phases was introduced [49].

The degree of functionalization was investigated in a previous study involving graphene flakes functionalized using the same reaction mechanism and parameters —with a closely similar azomethine ylide— which demonstrated a functionalization efficiency of approximately one ylide per ~170 carbon atoms, as estimated by X-ray photoelectron spectroscopy (XPS) analysis [31]. Therefore, it is reasonable to assume a comparable functionalization rate in this case as well.

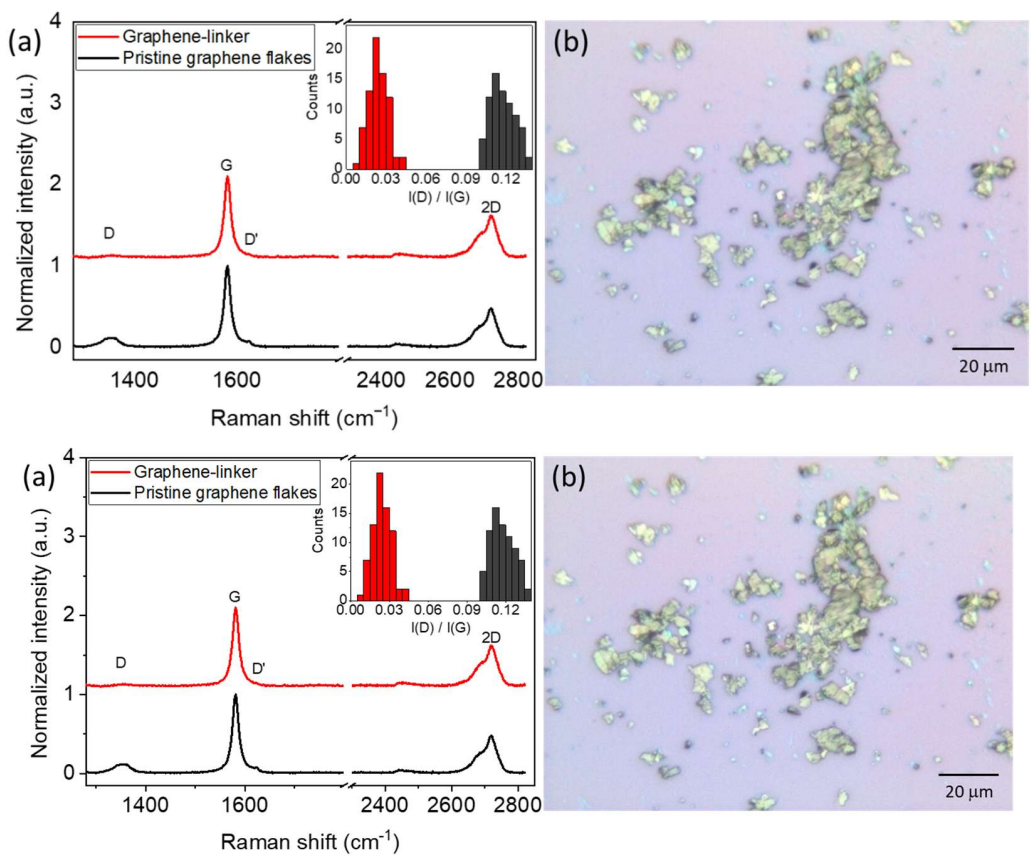

**Figure 3.** (**a**) Representative normalized Raman spectra acquired from pristine and linker-functionalized graphene flakes on Si/SiO$_2$. The inset shows a histogram of the I(D)/I(G) peak intensity ratio extracted from the Raman spectra (statistical analysis of 75 spectra acquired from different flakes). Histogram colors are used according to the colors of the curves shown in panel (**a**). (**b**) Optical bright field microscopy image (100× objective) of functionalized graphene flakes drop-casted onto a silica substrate for Raman measurements.

*3.3. Optical Characterization of Graphene Flakes Functionalized with the Fluorophores*

After the deprotection of the amine group and the reaction with the fluorophore Atto 465 NHS, photoluminescence (PL) spectroscopy was employed to confirm the successful functionalization of the flakes.

Frequently, for many compounds such as carbon nanomaterials, auto-fluorescence arises upon excitation in the green spectrum range [36]. For this reason, the optical characterization of a reference sample consisting of pristine graphene flakes was acquired. As demonstrated in Figure 4a, the PL spectrum of the pristine graphene flakes shows a negligible intrinsic fluorescence compared to the emission of the Atto 465-labeled graphene flakes, which is characterized by a broad band centered at about 520 nm, typical of this fluorophore (see Figure

S3 in the Supporting Information). In addition to the fluorescence, using a laser power of 6 mJ $\mu m^{-2}$, the bands associated with electronic transitions involving graphene are also evident at ~505 nm, 511 nm, and 543 nm. The corresponding Raman shifts are calculated considering the excitation wavelength used (473 nm) and yield values of approximately 1340 cm$^{-1}$, 1572 cm$^{-1}$, and 2725 cm$^{-1}$, thus ascribable to D, G, and 2D graphene bands, respectively. Furthermore, it is noteworthy that for laser energies exceeding ~12 mJ $\mu m^{-2}$, photo-bleaching was observed due to chemical damage and covalent modification [53].

To exclude both a possible signal arising from the graphene-linker system and the possibility that the fluorescent dye bonds directly to the graphene flakes, two more samples were prepared and optically characterized. One sample consisted of graphene flakes functionalized only with the linker molecule (performing only the first step of functionalization), and the second was obtained by directly performing the last reaction step between the pristine graphene flakes (without linker) and the fluorophore Atto 465 NHS under the conditions described in the Experimental Section. The PL spectra of the samples are presented in Figure 4b. Graphene flakes functionalized solely with the linker exhibit a PL spectrum similar to that of the pristine sample, showing no significant fluorescence signal (see Figure 4b). Also, the sample prepared without the initial cycloaddition step shows no detectable PL bands. This strongly suggests that no direct interaction occurs between the graphene flakes and the fluorophore, further confirming the absence of physisorption of the dye molecule onto the graphene surface.

After confirming that the PL signal was only observed when the fluorophore was chemically bound to graphene via the linker molecule, the Atto 465-labeled graphene flakes sample was further characterized using a confocal fluorescent microscope. At low magnification (5× objective), when the flakes on Si/SiO$_2$ are irradiated with blue light, they appear as uniformly distributed bright spots emitting light, as shown in Figure S4. Under excitation at 470 nm, which corresponds to the absorption maximum for the fluorophore, the emission spectra show a broad curve (FWHM ~120 nm) with a maximum at ~510 nm, as expected for this fluorophore (Figure 4c) [34]. This fluorescence profile reveals that the molecules do not undergo evident modifications upon graphene linkage. Compared to the functionalized sample, the pristine graphene flakes show a negligible emission due to a mild autofluorescence phenomenon at the set wavelength (Figure 4c). In addition, analysis of the fluorescent image (inset of Figure 4c) and the spectra in Figure S5 clearly indicates that the luminescence intensity—corresponding to the degree of surface functionalization—is relatively uniform across the graphene flakes. However, variations in luminescence intensity are observed across different ROIs, likely due to the random distribution of defects. While defects are expected to be more concentrated at the edges of graphene flakes [54], where covalent bonds readily form due to higher reactivity [55], the applied functionalization method effectively achieves a uniform derivatization of both the edges and the basal plane. Such homogeneous modification ensures consistent chemical interaction throughout the flakes, a critical factor for reproducibility and the optimization of material properties in various applications. Furthermore, the observed homogeneity in the fluorescence intensity distribution closely matches the nitrogen atom distribution determined in our prior work using energy-dispersive X-ray spectroscopy (EDX) following the grafting of a similar ylide onto graphene [31].

Along with the emission spectra, FLIM images of the graphene flakes were acquired to investigate the decay curves and the fluorescence lifetime. Due to the significant difference in fluorescence intensities observed between the functionalized and control samples, the curves were normalized for easier comparison. To further characterize the photophysical properties of the samples, the time-resolved room temperature decay curves of both the pure fluorophore and the Atto 465-labeled graphene samples at the same excitation wavelength of 470 nm were investigated, and the corresponding measurement results are presented in Figure 4d. The distinct decay profiles observed for these two emissions reveal different excited state dynamics, highlighting the different photophysical behavior of the fluorophore when interacting with graphene. The pure fluorophore Atto 465 dissolved in chloroform displays, as expected, a mono-exponential decay with a lifetime of ~3.5 ns [56,57]. On the contrary, Atto 465-labeled graphene flakes reveal a decay curve which can be fitted by a double exponential function, with estimated average lifetimes of $\tau_1 = 0.8$ ns, and $\tau_2 = 2.9$ ns (refer to Table S1 for further details).

Solutions of free fluorophores typically undergo a first-order decay process, governed by a single dominant relaxation pathway, leading to a single-exponential fluorescence decay. However, when these fluorophores are anchored to surfaces such as graphene or other substrates, their fluorescence decay behavior can change significantly. Multi-exponential decay patterns are commonly observed for fluorophores attached to molecules like nucleosides or cellular membranes [58,59], indicating that the fluorophores undergo multiple decay pathways due to various interactions with their environment.

In the case of Atto 465 anchored to graphene, the observed double-exponential decay curves can be attributed to multiple energy transfer processes occurring at the dye-substrate interface, including energy transfer to the graphene's conjugated π-system and charge interactions with the substrate [60]. This introduces a dual decay process: one representing rapid energy transfer to graphene and another corresponding to the intrinsic fluorescence decay of the fluorophore itself. Thus, the attachment to graphene alters the fluorescence dynamics by adding new decay channels, which reflects the complexity of interactions at the dye-substrate interface.

Moreover, the substrate can impose physical constraints on the fluorophores, affecting their rotational or conformational freedom, thereby influencing their radiative and non-radiative relaxation rates. The fluorophores' immobilization in a more rigid environment typically reduces non-radiative decay pathways, leading to an increase in fluorescence lifetime [61]. These micro-environmental factors contribute to the observed multi-exponential decay patterns, which reflect the combined effects of energy transfer, charge interactions, and molecular constraints on the fluorophores. In this regard, we observed that the lifetime values also could vary slightly considering different ROIs on the sample. For this reason, the lifetimes reported for the functionalized graphene samples are average values calculated considering several areas in the FLIM images.

Having confirmed that graphene flakes can indeed be labeled with dyes using the above-described procedure, we continued to label and characterize graphene flakes with other fluorescent dyes. In Figure 5, panel a, the optical characterization and fluorescence imaging of graphene flakes-Atto 425 NHS are shown. The emission spectrum of the sample was acquired using a confocal microscope under a 440 nm excitation wavelength (absorption maximum) [29]. The acquired spectrum, reported in Figure 5a, shows an emission curve with a maximum around 485 nm, comparable with the values reported in the literature for this fluorophore. As already observed under the previous excitation wavelength, negligible fluorescence was detected from the pristine sample.

To definitely exclude any possible interference due to autofluorescence phenomena from the samples, in a further experiment, the graphene flakes were functionalized with a fluorophore emitting in the red spectrum range, namely, Atto 633 NHS. As displayed in Figure 5b, upon excitation at 640 nm, the sample shows a fluorescence spectrum peaking around 652 nm, thus matching the literature data [29], while almost no fluorescence signal was detected from the control sample (pristine graphene flakes).

Once again, fluorescence imaging of these two samples (insets of Figure 5a,b) reveals a well-distributed functionalization across the entire surface, highlighting the uniformity of the surface functionalization. This observation further validates the effectiveness of the proposed process, confirming the consistent spatial distribution of molecules and highlighting the reliability of the applied modification strategy.

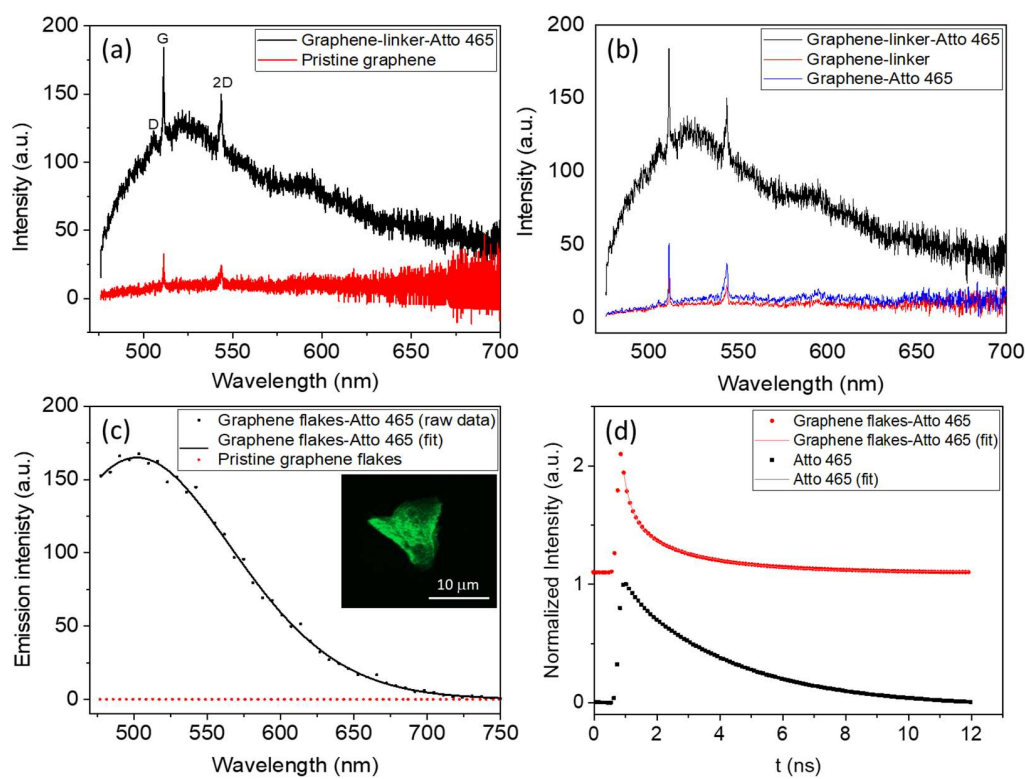

**Figure 4.** (**a**) PL spectra acquired at 473 nm excitation for Atto 465-labeled graphene flakes and the pristine sample. (**b**) PL spectra of functionalized graphene flakes (black curve) and control samples: graphene flakes functionalized with only the linker molecule (red line) and graphene flakes not subjected to the first reaction step (no linker), merely kept in contact with the fluorophore (blue line). (**c**) Normalized fluorescence spectra acquired with a confocal microscope under excitation at 470 nm, showing the graphene flakes before and after functionalization with Atto 465 NHS. The inset in panel (**c**) shows a fluorescence image of the flakes. (**d**) Normalized fluorescence decay curves of the sample functionalized with Atto 465 compared to the pure fluorophores in chloroform (3 μg μL$^{-1}$), obtained from FLIM images.

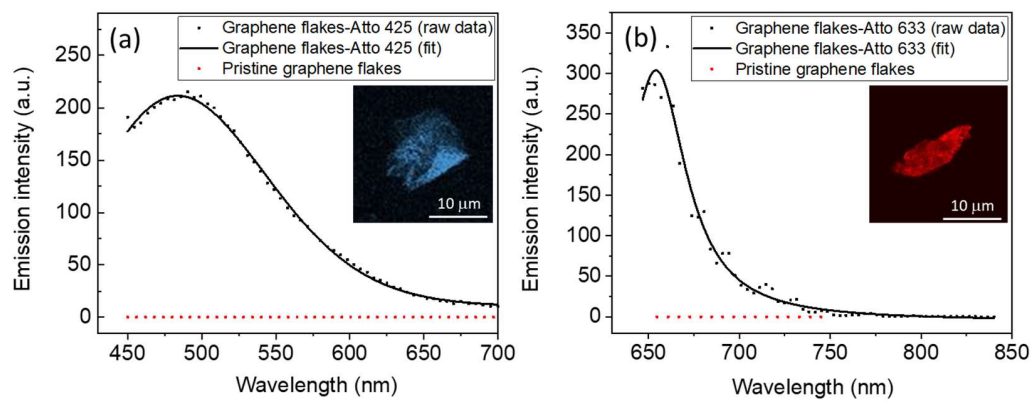

**Figure 5.** Fluorescence spectra of graphene flakes functionalized with (**a**) Atto 425 NHS and (**b**) Atto 633 NHS compared with the blank signal. The data were acquired with a confocal microscope under excitation at 440 nm and 640 nm, respectively. The insets show representative fluorescence images of the flakes.

Consistent with the previous results, graphene flakes labeled with both Atto 425 and Atto 633 exhibited decay curves similar to those observed using Atto 465 (refer to Figure S6, panels a, b). The lifetime values obtained for each functionalized sample, as well as for each free fluorophore in solution, are reported in Table S1 in the Supporting Information.

Besides the possibility to use the linker molecule as a platform to label graphene with a variety of molecules, another compelling aspect of the graphene samples investigated is that after storage in ambient conditions, protected from direct light exposure, they were found to be stable for over a year, i.e., over time they retain the ability to emit upon absorption at a proper wavelength.

## 4. Conclusions

In this work, we presented a simple mechanism to successfully functionalize low-defect graphene flakes with fluorescent dyes through the linkage of an organic molecule. The functionalization reaction was carried out in three steps involving (I) the 1,3-dipolar cycloaddition of a custom-synthesized azomethine ylide to graphene, (II) deprotection of a terminal amine group, and (III) chemical binding of a fluorophore to the linker molecule using NHS ester derivatization. The fluorescence spectra acquired from the samples labeled with three different fluorophores (Atto 425, Atto 465, and Atto 633 NHS ester) showed the typical emission curves of the dyes, demonstrating the effectiveness of the procedure. This proof-of-concept experiment sets the stage for exploiting linker molecules attached to graphene in the development of platforms for various applications, thanks to the possibility to subsequently attach molecules or functional groups to the linkers. The potential to functionalize low-defect graphene systems through this approach enables the design of customized devices, where the unique properties of graphene can be synergistically combined with those of other molecules, nanoparticles, or quantum dots by a straightforward and easily controllable mechanism.

**Supplementary Materials:** Figure S1: UHPLC-MS analysis of the linker (a) before and (b) after its purification (retention time = 11.5 ± 0.1 min). (c) Mass spectrum of the product fragmentation pattern of ions and molecular structure. Figure S2: FT-IR spectra (linear scale) acquired for the cleavage solution (phosphoric acid/$CH_2Cl_2$/$H_2O$, black line) and at the end of the reaction (t = 6.5 h, red line) demonstrating the successful deprotection reaction. The spectra were acquired after complete evaporation of $CH_2Cl_2$. Figure S3: PL spectra acquired with at 473 nm excitation: comparison between the normalized spectra of Atto 465 NHS in DMSO (1 μg μL$^{-1}$) drop-casted on a glass slice (red line) and graphene flakes functionalized with the same fluorophore (black line). Figure S4: Optical image of the graphene flakes functionalized with Atto 465 under blue LED light illumination, acquired with a 5× objective. The sample is fully covered by functionalized flakes, except for the lower right corner of the image, where a scratch has removed the flakes, and no fluorescence is detected. Figure S5: (a) Fluorescence spectra of Atto 465-labeled graphene acquired with a confocal microscope in two regions of interest (ROI) shown in panel (b). Excitation wavelength: 470 nm. The luminescence intensity, indicating the degree of functionalization, is not homogeneous across the flakes. This variation is attributed to the typical defect distribution, which is higher at the edges of the graphene flakes. However, it is noteworthy that functionalization also extends to the basal planes, contributing to the overall luminescence pattern. Figure S6: Normalized fluorescence decay with fitting curves of the samples functionalized with (a) Atto 425 and (b) Atto 633 compared to those typical of the pure fluorophores in chloroform (3 μg μL$^{-1}$), obtained from FLIM images. Table S1: Lifetimes obtained for graphene flakes functionalized with Atto 465, Atto 425, and Atto 633 and for the respective pure fluorophores in chloroform (3 μg μL$^{-1}$). Average value ± standard deviation from five measurements.

**Acknowledgments:** The authors gratefully acknowledge BeDimensional S.p.A. for providing graphene powder, Filippo Fabbri from Istituto Nanoscienze–CNR for his help with PL measurements, and Camilla Coletti from Istituto Italiano di Tecnologia (CNI @ NEST) for giving access to the Raman instrumentation. Y.V. acknowledges FSE-REACT-EU Programma Operativo Nazionale "Ricerca e Innovazione" 2014–2020 for the support.

# References

1. Novoselov, K.S.; Geim, A.K.; Morozov, S.V.; Jiang, D.; Zhang, Y.; Dubonos, S.V.; Grigorieva, I.V.; Firsov, A.A. Electric Field Effect in Atomically Thin Carbon Films. *Science* **2004**, *306*, 666–669. https://doi.org/10.1126/science.1102896.
2. Soldano, C.; Mahmood, A.; Dujardin, E. Production, properties and potential of graphene. *Carbon* **2010**, *48*, 2127–2150. https://doi.org/10.1016/j.carbon.2010.01.058.
3. Novoselov, K.S.; Morozov, S.V.; Mohinddin, T.M.G.; Ponomarenko, L.A.; Elias, D.C.; Yang, R.; Barbolina, I.I.; Blake, P.; Booth, T.J.; Jiang, D.; et al. Electronic properties of graphene. *Phys. Status Solidi Basic Res.* **2007**, *244*, 4106–4111. https://doi.org/10.1002/pssb.200776208.
4. Sun, X.; Huang, C.; Wang, L.; Liang, L.; Cheng, Y.; Fei, W.; Li, Y. Recent Progress in Graphene/Polymer Nanocomposites. *Adv. Mater.* **2021**, *33*, 2001105. https://doi.org/10.1002/adma.202001105.
5. Ke, Q.; Wang, J. Graphene-based materials for supercapacitor electrodes—A review. *J. Mater.* **2016**, *2*, 37–54. https://doi.org/10.1016/j.jmat.2016.01.001.
6. Goenka, S.; Sant, V.; Sant, S. Graphene-based nanomaterials for drug delivery and tissue engineering. *J. Control. Release* **2014**, *173*, 75–88. https://doi.org/10.1016/j.jconrel.2013.10.017.
7. Wang, X.; Zhi, L.; Müllen, K. Transparent, conductive graphene electrodes for dye-sensitized solar cells. *Nano Lett.* **2008**, *8*, 323–327. https://doi.org/10.1021/nl072838r.

8. Ji, Y.; Lee, S.; Cho, B.; Song, S.; Lee, T. Flexible organic memory devices with multilayer graphene electrodes. *ACS Nano* **2011**, *5*, 5995–6000. https://doi.org/10.1021/nn201770s.
9. Lai, S.; Vlamidis, Y.; Mishra, N.; Cosseddu, P.; Mišeikis, V.; Ricci, P.C.; Voliani, V.; Coletti, C.; Bonfiglio, A. A Flexible, Transparent Chemosensor Integrating an Inkjet-Printed Organic Field-Effect Transistor and a Non-Covalently Functionalized Graphene Electrode. *Adv. Mater. Technol*. **2021**, *6*, 2100481. https://doi.org/10.1002/admt.202100481.
10. Alwarappan, S.; Erdem, A.; Liu, C.; Li, C.Z. Probing the electrochemical properties of graphene nanosheets for biosensing applications. *J. Phys. Chem. C* **2009**, *113*, 8853–8857. https://doi.org/10.1021/jp9010313.
11. Mao, H.Y.; Lu, Y.H.; Lin, J.D.; Zhong, S.; Wee, A.T.S.; Chen, W. Manipulating the electronic and chemical properties of graphene via molecular functionalization. *Prog. Surf. Sci*. **2013**, *88*, 132–159. https://doi.org/10.1016/j.progsurf.2013.02.001.
12. Guo, S.; Garaj, S.; Bianco, A.; Ménard-Moyon, C. Controlling covalent chemistry on graphene oxide. *Nat. Rev. Phys.* **2022**, *4*, 247–262. https://doi.org/10.1038/s42254-022-00422-w.
13. Rabchinskii, M.K.; Ryzhkov, S.A.; Besedina, N.A.; Brzhezinskaya, M.; Malkov, M.N.; Stolyarova, D.Y.; Arutyunyan, A.F.; Struchkov, N.S.; Saveliev, S.D.; Diankin, I.D.; et al. Guiding graphene derivatization for covalent immobilization of aptamers. *Carbon* **2022**, *196*, 264–279. https://doi.org/10.1016/j.carbon.2022.04.072.
14. Rabchinskii, M.K.; Sysoev, V.V.; Varezhnikov, A.S.; Solomatin, M.A.; Struchkov, N.S.; Stolyarova, D.Y.; Ryzhkov, S.A.; Antonov, G.A.; Gabrelian, V.S.; Cherviakova, P.D.; et al. Toward On-Chip Multisensor Arrays for Selective Methanol and Ethanol Detection at Room Temperature: Capitalizing the Graphene Carbonylation. *ACS Appl. Mater. Interfaces* **2023**, *15*, 23, 28370–28386. https://doi.org/10.1021/acsami.3c02833.
15. Maio, A.; Pibiri, I.; Morreale, M.; La Mantia, F.P.; Scaffaro, R. An Overview of Functionalized Graphene Nanomaterials for Advanced Applications. *Nanomaterials* **2021**, *11*, 1717. https://doi.org/10.3390/nano11071717.
16. Boukhvalov, D.W.; Katsnelson, M.I. Chemical Functionalization of Graphene with Defects. *Nano Lett*. **2008**, *8*, 4373–4379. https://doi.org/10.1021/nl802234n.
17. Ioniţă, M.; Vlăsceanu, G.M.; Watzlawek, A.A.; Voicu, S.I.; Burns, J.S.; Iovu, H. Graphene and functionalized graphene: Extraordinary prospects for nanobiocomposite materials. *Compos. Part B Eng.* **2017**, *121*, 34–57. https://doi.org/10.1016/j.compositesb.2017.03.031.
18. Shan, C.; Yang, H.; Han, D.; Zhang, Q.; Ivaska, A.; Niu, L. Water-soluble graphene covalently functionalized by biocompatible poly-L-lysine. *Langmuir* **2009**, *25*, 12030–12033. https://doi.org/10.1021/la903265p.
19. Hu, H.; Allan, C.C.K.; Li, J.; Kong, Y.; Wang, X.; Xin, J.H.; Hu, H. Multifunctional organically modified graphene with super-hydrophobicity. *Nano Res*. **2014**, *7*, 418–433. https://doi.org/10.1007/s12274-014-0408-0.
20. Loh, K.P.; Bao, Q.; Ang, P.K.; Yang, J. The chemistry of graphene. *J. Mater. Chem*. **2010**, *20*, 2277–2289. https://doi.org/10.1039/b920539j.
21. Yang, G.-h.; Bao, D.-d., Liu, H.; Zhang, D.-q.; Wang, N.; Li, H.-t. Functionalization of Graphene and Applications of the Derivatives. *J. Inorg. Organomet. Polym. Mater*. **2017**, *27*, 1129–1141. https://doi.org/10.1007/s10904-017-0597-6.
22. Criado, A.; Melchionna, M.; Marchesan, S.; Prato, M. The Covalent Functionalization of Graphene on Substrates. *Angew. Chem. Int. Ed*. **2015**, *54*, 10734–10750. https://doi.org/10.1002/anie.201501473.

23. Wetzl, C.; Silvestri, A.; Garrido, M.; Hou, H.L.; Criado, A.; Prato, M. The Covalent Functionalization of Surface-Supported Graphene: An Update. *Angew. Chem. Int. Ed.* **2023**, *62*, e202212857. https://doi.org/10.1002/anie.202212857.
24. Choi, J.; Kim, K.J.; Kim, B.; Lee, H.; Kim, S. Covalent functionalization of epitaxial graphene by azidotrimethylsilane. *J. Phys. Chem. C* **2009**, *113*, 9433–9435. https://doi.org/10.1021/jp9010444.
25. Kim, K.H.; Seo, S.E.; Park, S.J.; Kim, J.; Park, C.S.; Le, T.H.; Lee, C.S.; Kim, Y.K.; Kim, H.Y.; Jun, S.; et al. N-Heterocyclic Carbene–Graphene Nanotransistor Based on Covalent Bond for Ultrastable Biosensors. *Adv. Funct. Mater.* **2024**, *34*, 2310377. https://doi.org/10.1002/adfm.202310377.
26. Sulleiro, M.V.; Quiroga, S.; Peña, D.; Pérez, D.; Guitián, E.; Criado, A.; Prato, M. Microwave-induced covalent functionalization of few-layer graphene with arynes under solvent-free conditions. *Chem. Commun.* **2018**, *54*, 2086–2089. https://doi.org/10.1039/c7cc08676h.
27. Georgakilas, V.; Bourlinos, A.B.; Zboril, R.; Steriotis, T.A.; Dallas, P.; Stubos, A.K.; Trapalis, C. Organic functionalisation of graphenes. *Chem. Commun.* **2010**, *46*, 1766–1768. https://doi.org/10.1039/b922081j.
28. Wu, S.; Sun, W.; Zhang, D.; Shu, L.; Wu, H.; Xu, J.; Lao, X. 1,3-Dipolar cycloaddition of several azomethine ylides to [60]fullerene: Synthesis of derivatives of 2′,5′-dihydro-1′H-pyrrolo[3′,4′:1,2][60]fullerene 1. *J. Chem. Soc., Perkin Trans.* **1998**, *1*, 1733–1738. https://doi.org/10.1039/A705962K.
29. Bayazit, M.K.; Coleman, K.S. Fluorescent single-walled carbon nanotubes following the 1,3-dipolar cycloaddition of pyridinium ylides. *J. Am. Chem. Soc.* **2009**, *131*, 10670–10676. https://doi.org/10.1021/ja903712f.
30. Filippone, S.; Maroto, E.E.; Martín-Domenech, Á.; Suarez, M.; Martín, N. An efficient approach to chiral fullerene derivatives by catalytic enantioselective 1,3-dipolar cycloadditions. *Nat. Chem.* **2009**, *1*, 578–582. https://doi.org/10.1038/nchem.361.
31. Basta, L.; Moscardini, A.; Fabbri, F.; Bellucci, L.; Tozzini, V.; Rubini, S.; Griesi, A.; Gemmi, M.; Heun, S.; Veronesi, S. Covalent organic functionalization of graphene nanosheets and reduced graphene oxide via 1,3-dipolar cycloaddition of azomethine ylide. *Nanoscale Adv.* **2021**, *3*, 5841–5852. https://doi.org/10.1039/d1na00335f.
32. Basta, L.; Bianco, F.; Moscardini, A.; Fabbri, F.; Bellucci, L.; Tozzini, V.; Heun, S.; Veronesi, S. Deterministic Covalent Organic Functionalization of Monolayer Graphene with 1,3-Dipolar Cycloaddition Via High Resolution Surface Engineering. *J. Mater. Chem. C* **2022**, *11*, 2630–2639. https://doi.org/10.2139/ssrn.4039980.
33. Lukumbuzya, M.; Schmid, M.; Pjevac, P.; Daims, H. A multicolor fluorescence in situ hybridization approach using an extended set of fluorophores to visualize microorganisms. *Front. Microbiol.* **2019**, *10*, 1383. https://doi.org/10.3389/fmicb.2019.01383.
34. Dodge, J.T.; Doyle, A.D.; Costa-da-Silva, A.C.; Hogden, C.T.; Mezey, E.; Mays, J.W. Atto 465 Derivative Is a Nuclear Stain with Unique Excitation and Emission Spectra Useful for Multiplex Immunofluorescence Histochemistry. *J. Histochem. Cytochem.* **2022**, *70*, 211–223. https://doi.org/10.1369/00221554211064942.
35. Hübner, K.; Joshi, H.; Aksimentiev, A.; Stefani, F.D.; Tinnefeld, P.; Acuna, G.P. Determining the In-Plane Orientation and Binding Mode of Single Fluorescent Dyes in DNA Origami Structures. *ACS Nano* **2021**, *15*, 5109–5117. https://doi.org/10.1021/acsnano.0c10259.

36. Krasley, A.T.; Li, E.; Galeana, J.M.; Bulumulla, C.; Beyene, A.G.; Demirer, G.S. Carbon Nanomaterial Fluorescent Probes and Their Biological Applications. *Chem. Rev.* **2024**, *124*, 3085–3185. https://doi.org/10.1021/acs.chemrev.3c00581.
37. Peng, C.; Hu, W.; Zhou, Y.; Fan, C.; Huang, Q. Intracellular imaging with a graphene-based fluorescent probe. *Small* **2010**, *6*, 1686–1692. https://doi.org/10.1002/smll.201000560.
38. Halbig, C.E.; Lasch, R.; Krüll, J.; Pirzer, A.S.; Wang, Z.; Kirchhof, J.N.; Bolotin, K.I.; Heinrich, M.R.; Eigler, S. Selective Functionalization of Graphene at Defect-Activated Sites by Arylazocarboxylic tert-Butyl Esters. *Angew. Chem. Int. Ed.* **2019**, *58*, 3599–3603. https://doi.org/10.1002/anie.201811192.
39. Lim, C.Y.; Owens, N.A.; Wampler, R.D.; Ying, Y.; Granger, J.H.; Porter, M.D.; Takahashi, M.; Shimazu, K. Succinimidyl ester surface chemistry: Implications of the competition between aminolysis and hydrolysis on covalent protein immobilization. *Langmuir* **2014**, *30*, 12868–12878. https://doi.org/10.1021/la503439g.
40. Guo, Z.; Chakraborty, S.; Monikh, F.A.; Varsou, D.-D.; Chetwynd, A.J.; Afantitis, A.; Lynch, I.; Zhang, P. Surface functionalization of graphene-based materials: Biological behavior, toxicology, and safe-by-design aspects. *Adv. Biol.* **2021**, *5*, 2100637. https://doi.org/10.1002/adbi.202100637.
41. Isidro-Llobet, A.; Álvarez, M.; Albericio, F. Amino Acid-Protecting Groups. *Chem. Rev.* **2009**, *109*, 2455–2504. https://doi.org/10.1021/cr800323s.
42. Tang, Y.; Ma, J.; Imler, G.H.; Parrish, D.A.; Shreeve, J.M. Versatile functionalization of 3,5-diamino-4-nitropyrazole for promising insensitive energetic compounds. *Dalt. Trans.* **2019**, *48*, 14490–14496. https://doi.org/10.1039/c9dt03138c.
43. Bütikofer, A.; Chen, P. Zwitterionic Halido Cyclopentadienone Iron Complexes and Their Catalytic Performance in Hydrogenation Reactions. *Inorg. Chem.* **2023**, *62*, 4188–4196. https://doi.org/10.1021/acs.inorgchem.2c04298.
44. Lee, D.W.; Ha, H.J.; Lee, W.K. Selective mono-BOC protection of diamines. *Synth. Commun.* **2007**, *37*, 737–742. https://doi.org/10.1080/00397910601131403.
45. George, N.; Ofori, S.; Parkin, S.; Awuah, S.G. Mild deprotection of the: N-tert-butyloxycarbonyl (N-Boc) group using oxalyl chloride. *RSC Adv.* **2020**, *10*, 24017–24026. https://doi.org/10.1039/d0ra04110f.
46. Fattahi, N.; Ayubi, M.; Ramazani, A. Amidation and esterification of carboxylic acids with amines and phenols by N,N′-diisopropylcarbodiimide: A new approach for amide and ester bond formation in water. *Tetrahedron* **2018**, *74*, 4351–4356. https://doi.org/10.1016/j.tet.2018.06.064.
47. Datta, R.; Heaster, T.M.; Sharick, J.T.; Gillette, A.A.; Skala, M.C. Fluorescence lifetime imaging microscopy: Fundamentals and advances in instrumentation, analysis, and applications. *J. Biomed. Opt.* **2020**, *25*, 071203. https://doi.org/10.1117/1.JBO.25.7.071203.
48. Del Rio Castillo, A.E.; Pellegrini, V.; Ansaldo, A.; Ricciardella, F.; Sun, H.; Marasco, L.; Buha, J.; Dang, Z.; Gagliani, L.; Lago, E.; et al. High-yield production of 2D crystals by wet-jet milling. *Mater. Horizons* **2018**, *5*, 890–904. https://doi.org/10.1039/c8mh00487k.
49. Ferrari, A.C.; Meyer, J.C.; Scardaci, V.; Casiraghi, C.; Lazzeri, M.; Mauri, F.; Piscanec, S.; Jiang, D.; Novoselov, K.S.; Roth, S.; et al. Raman Spectrum of Graphene and Graphene Layers. *Phys. Rev. Lett.* **2006**, *97*, 187401. https://doi.org/10.1103/PhysRevLett.97.187401.
50. Jorio, A.; Ferreira, E.H.M.; Stavale, F.; Achete, C.A.; Capaz, R.B.; Moutinho, M.V.O.; Lombardo, A.; Kulmala, T.S.; Ferrari, A.C. Quantifying Defects in Graphene via Raman

Spectroscopy at Different Excitation Energies. *Nano Lett*. **2011**, *12*, 3190–3196. https://doi.org/10.1021/nl201432g.

51. Eckmann, A.; Felten, A.; Mishchenko, A.; Britnell, L.; Krupke, R.; Novoselov, K.S.; Casiraghi, C. Probing the nature of defects in graphene by Raman spectroscopy. *Nano Lett*. **2012**, *12*, 3925–3930. https://doi.org/10.1021/nl300901a.
52. Eckmann, A.; Felten, A.; Verzhbitskiy, I.; Davey, R.; Casiraghi, C. Raman study on defective graphene: Effect of the excitation energy , type , and amount of defects. *Phys. Rev. B* **2013**, *88*, 035426. https://doi.org/10.1103/PhysRevB.88.035426.
53. MacHáň, R.; Hof, M.; Chernovets, T.; Zhmak, M.N.; Ovchinnikova, T.V.; Sýkora, J. Formation of arenicin-1 microdomains in bilayers and their specific lipid interaction revealed by Z-scan FCS. *Anal. Bioanal. Chem*. **2011**, *399*, 3547–3554. https://doi.org/10.1007/s00216-011-4694-z.
54. Sethurajaperumal, A.; Ravichandran, V.; Merenkov, I.; Ostrikov (Ken), Varrla, E. Delamination and defects in graphene nanosheets exfoliated from 3D precursors. *Carbon* **2023**, *213*, 118306. https://doi.org/10.1016/j.carbon.2023.118306.
55. Acik, M.; Chabal, Y.J. Nature of graphene edges: A review. *Jpn. J. Appl. Phys*. **2011**, *50*, 070101. https://doi.org/10.1143/JJAP.50.070101.
56. Chmyrov, A.; Arden-Jacob, J.; Zilles, A.; Drexhage, K.H.; Widengren, J. Characterization of new fluorescent labels for ultra-high resolution microscopy. *Photochem. Photobiol. Sci*. **2008**, *7*, 1378–1385. https://doi.org/10.1039/b810991p.
57. Arden-Jacob, J.; Drexhage, K.H.; Druzhinin, S.I.; Ekimova, M.; Flender, O.; Lenzer, T.; Oum, K.; Scholz, M. Ultrafast photoinduced dynamics of the 3,6-diaminoacridinium derivative ATTO 465 in solution. *Phys. Chem. Chem. Phys*. **2013**, *15*, 1844–1853. https://doi.org/10.1039/c2cp43493h.
58. Hammler, D.; Marx, A.; Zumbusch, A. Fluorescence-Lifetime-Sensitive Probes for Monitoring ATP Cleavage. *Chem. Eur. J*. **2018**, *24*, 15329–15335. https://doi.org/10.1002/chem.201803234.
59. Chen, T.; Karedla, N.; Enderlein, J. Measuring sub-nanometer undulations at microsecond temporal resolution with metal- and graphene-induced energy transfer spectroscopy. *Nat. Commun*. **2024**, *15*, 1789. https://doi.org/10.1038/s41467-024-45822-x.
60. Xin, G.; Gong, S.; Kim, N.; Kim, J.; Hwang, W.; Nam, J.; Cho, Y.H.; Cho, S.M.; Chae, H.; Graphene oxide/N-methyl-2-pyrrolidone charge-transfer complexes for molecular detection. *Sens. Actuators B Chem*. **2013**, *176*, 81–85. https://doi.org/10.1016/j.snb.2012.09.003.
61. Berezin, M.Y.; Achilefu, S. Fluorescence lifetime measurements and biological imaging. *Chem. Rev*. **2010**, *110*, 2641–2684. https://doi.org/10.1021/cr900343z.

# Surface grafting of graphene flakes with fluorescent dyes: a tailored functionalization approach

*Ylea Vlamidis [1,2,\*], Carmela Marinelli [2], Aldo Moscardini [3], Paolo Faraci [3], Stefan Heun [1,\*],*

*Stefano Veronesi [1]*

[1] NEST, Istituto Nanoscienze–CNR and Scuola Normale Superiore, Piazza San Silvestro 12, 56127 Pisa, Italy

[2] Department of Physical Science, Earth, and Environment, University of Siena, via Roma 56, 53100 Siena, Italy

[3] Scuola Normale Superiore, Laboratorio NEST, Piazza San Silvestro 12, 56127 Pisa, Italy

*Corresponding Author. E-mail address: ylea.vlamidis@unisi.it

## 1. Analytical techniques and equipment

### 1.1 UHPLC-MS analysis of the linker molecule

The linker molecule was analyzed and purified with an Ultra High Performance Liquid Chromatography (UHPLC) system (Shimadzu Nexera). The analyte was identified through a diode-array detector using a wavelength of 227 nm and confirmed with a triple quadrupole mass spectrometer (AB Sciex 3200 QTRAP) equipped with an electrospray ionization (ESI) source.
For the analysis, a Kinetex EVO C18 5 μm column (3.0 × 150 mm) was employed. Eluent A was constituted by a 15 mM ammonium acetate buffer (pH 8.0), while eluent B was a mixture of acetonitrile:eluent A (95:5). The obtained product (~8 mg) was dissolved in a mixture of Eluent A:Eluent B (3:1). The following conditions were employed: flow rate 0.8 mL min$^{-1}$; gradient separation; first 3 minutes 10% of Eluent B, until 90% of B was reached after 15 minutes, this ratio was maintained for further 4 minutes.

The following mass spectrometer (MS) conditions were employed. Ionization mode: ESI; $N_2$ gas flow 25 mL min$^{-1}$; ion spray voltage: 5500 V; temperature: 120 °C; ion source gas 1 55.0 mL min$^{-1}$; ion source gas 2.60 mL min$^{-1}$; declustering potential: 40 V; entrance potential: 10 V; collision energy: 10 eV.

In Figure S1, the UHPLC-MS analysis of the linker (a) before and (b) after its purification are shown. Using the mix of eluents described before, the retention time of the product was 11.5 ± 0.1 min. The mass spectrum of the product fragmentation pattern of ions is reported in panel (c).

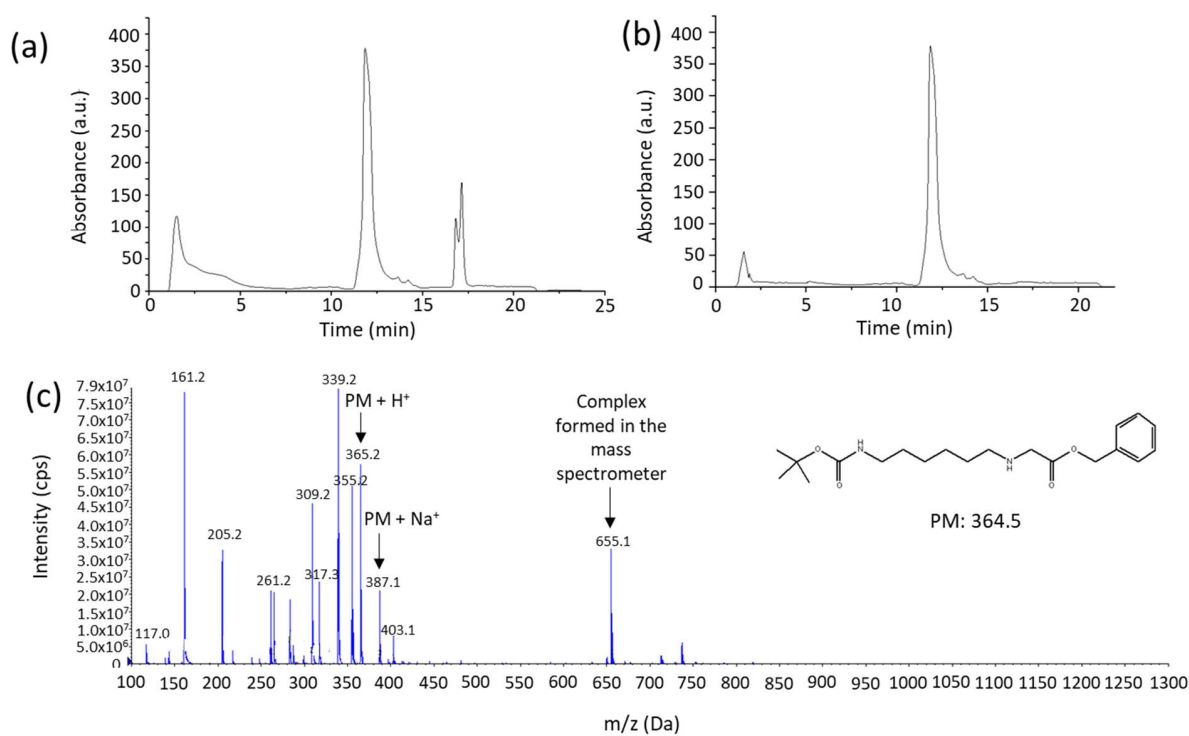

**Figure S1.** UHPLC-MS analysis of the linker (a) before and (b) after its purification (retention time = 11.5 ± 0.1 min). (c) Mass spectrum of the product fragmentation pattern of ions and molecular structure.

## 1.2 FT-IR spectra analysis

Fourier Transform-Infrared Spectroscopy (FT-IR) was performed with an Agilent Technologies, Cary 630 FT-IR Spectrometer, to confirm the Boc cleavage in the second step of the functionalization reaction. FT-IR spectra were acquired using Agilent MicroLab FT-IR software. The spectra were recorded between 4000 and 650 cm$^{-1}$, with a resolution of 4 cm$^{-1}$.

Figure S2 shows the FT-IR spectra of the cleavage solution before and after amine deprotection. The spectra were acquired after complete evaporation of the dichloromethane, to avoid the occurrence of strong absorption bands arising from the solvent.

In the spectra of the cleavage solution, the absorption broad bands around 3500 cm$^{-1}$ are the characteristic absorption peaks of water molecules or hydroxyl containing molecules (likes alcohols) [1]. The peaks between 1200 and 1300 cm$^{-1}$ are related to the P=O stretching vibrations of phosphoric acid [2]. The peaks near 1000 cm$^{-1}$ correspond to the symmetrical stretching of P-O-H, while the peak centered at ~730 cm$^{-1}$ represent the P-O-H bending mode [3].

The FT-IR spectrum of the product after Boc cleavage shows peaks between 3600 and 3300 cm$^{-1}$ which are due to O-H stretching vibrations.[1] The new strong bands emerging between 2850 and 2950 cm$^{-1}$ correspond to the aliphatic C-H stretching vibration[4] and suggest that the deprotection reaction is complete. The band at 1730 cm$^{-1}$ corresponds to the C=O stretch vibration of the amide, while the absorption bands between 1400 and 1300 cm$^{-1}$ are ascribable to C-H bending mode and O-H in-plane bending vibration [4,5]. The bands due to C-O stretching of tertiary alcohol are observable between 1000 and 1100 cm$^{-1}$[6]. Eventually, the band at ~710 cm$^{-1}$ is caused by the O-H out-of-plane bending vibration [7].

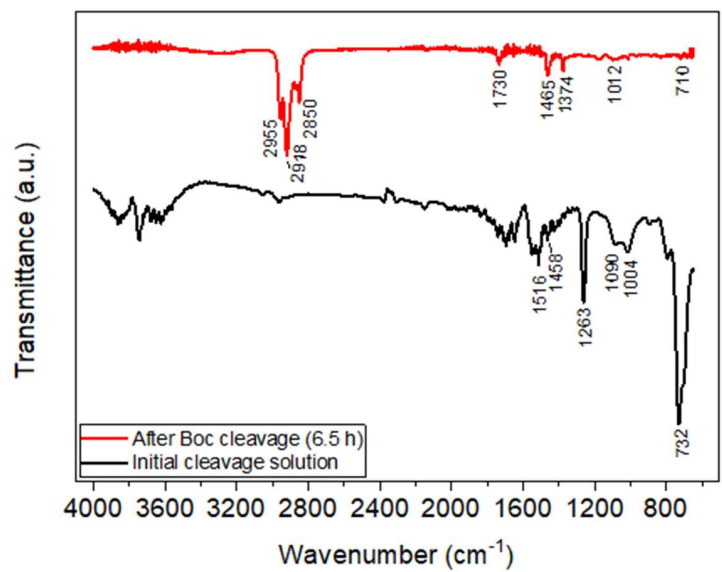

**Figure S2.** FT-IR spectra (linear scale) acquired for the cleavage solution (phosphoric acid /$CH_2Cl_2$/$H_2O$, black line) and at the end of the reaction (t = 6.5 h, red line) demonstrating the successful deprotection reaction. The spectra were acquired after complete evaporation of $CH_2Cl_2$.

## 2. Further fluorescence characterization of the graphene flakes labeled with fluorophores

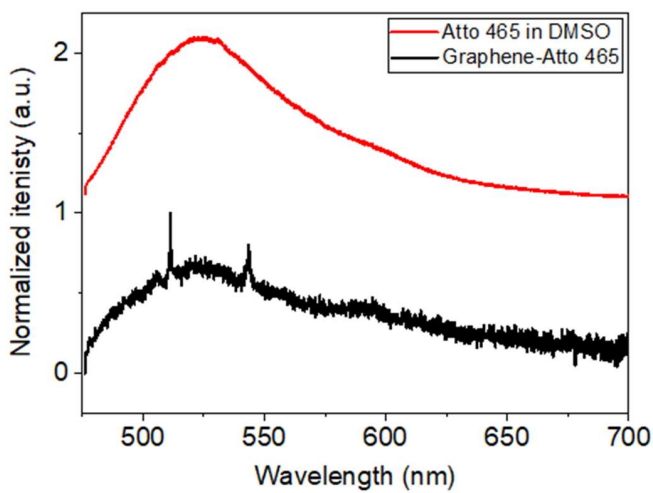

**Figure S3.** PL spectra acquired with at 473 nm excitation: comparison between the normalized spectra of Atto 465 NSH in DMSO (1 µg µL-1) drop-casted on a glass slice (red line) and graphene flakes functionalized with the same fluorophore (black line).

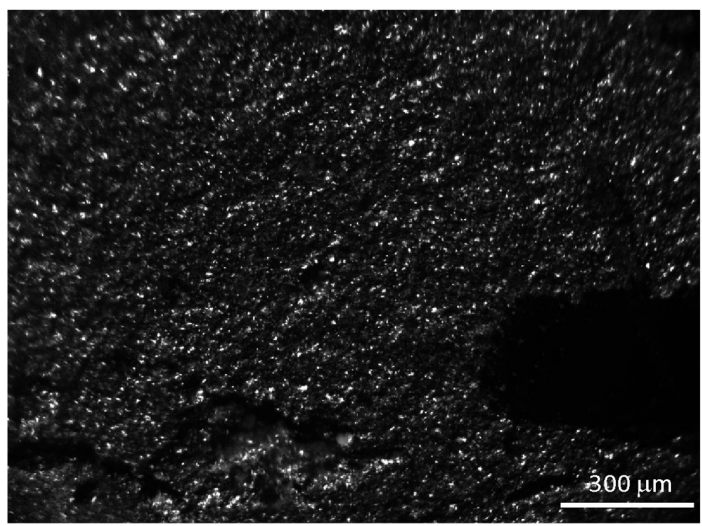

**Figure S4.** Optical image of the graphene flakes functionalized with Atto 465 under blue LED light illumination, acquired with a 5x objective. The sample is fully covered by functionalized flakes, except for the lower right corner of the image, where a scratch has removed the flakes and no fluorescence is detected.

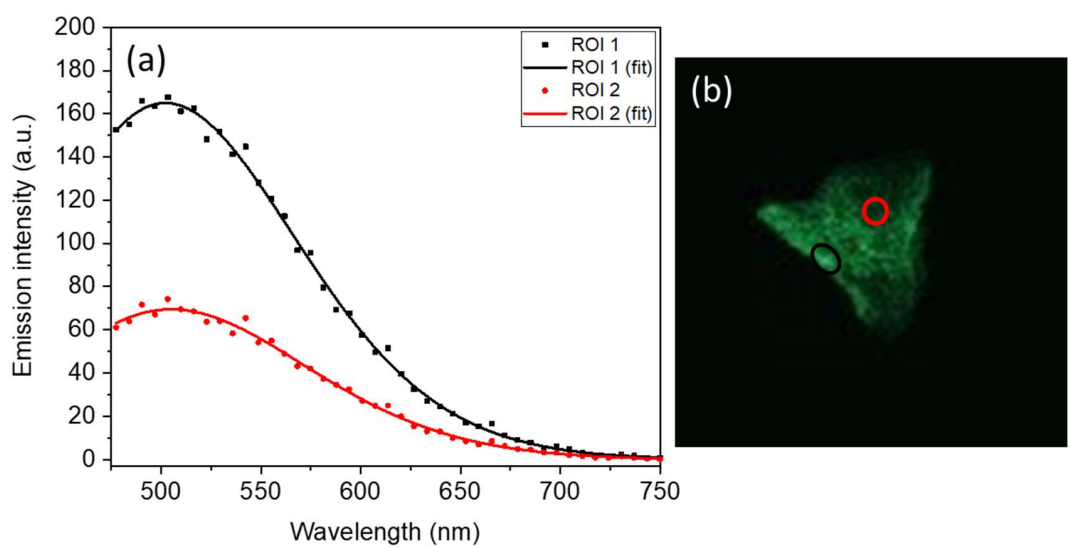

**Figure S5.** (a) Fluorescence spectra of Atto 465-labeled graphene acquired with a confocal microscope in two regions of interest (ROI) shown in panel (b). Excitation wavelength: 470 nm. The luminescence intensity, indicating the degree of functionalization, is not homogeneous across the flakes. This variation is attributed to the typical defect distribution, which is higher at the edges of the graphene flakes. However, it is noteworthy that functionalization also extends to the basal planes, contributing to the overall luminescence pattern.

## 2.1 Lifetime measurements

For the lifetime measurements of the graphene flakes functionalized with dyes, the total fluorescence emitted by the samples was collected, whereas the control samples consisted of the pure fluorophores dissolved in chloroform (3 µg µL-1). Calibration of the system was performed using the lifetime decay of a stock solution of fluorescein in ethanol (100 µM) diluted 1:500 in 0.1 M NaOH (pH =11). Excitation was performed at a wavelength of 488 nm, and fluorescence was collected in the range 400–570 nm.

Similarly to what already observed in the case of graphene flakes labeled with Atto 465, both graphene samples functionalized with Atto 425 and Atto 633 exhibited double exponential fluorescence decay curves, whereas typical first order reactions were observed for the free fluorophores in solution (refer to Figure S6). The average lifetimes obtained for each sample analyzed are reported in Table S1.

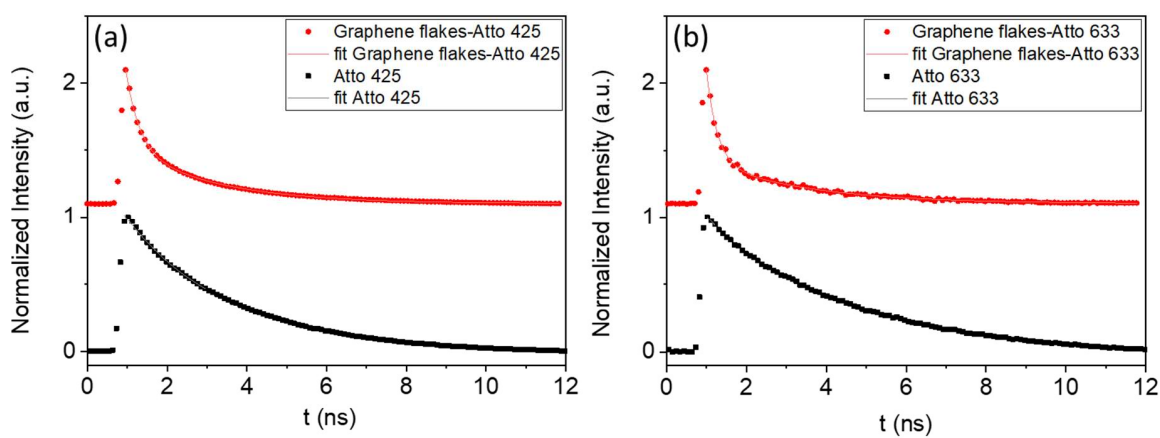

**Figure S6.** Normalized fluorescence decay with fitting curves of the samples functionalized with (a) Atto 425 and (b) Atto 633 compared to those typical of the pure fluorophores in chloroform (3 µg µL$^{-1}$), obtained from FLIM images.

**Table S1.** Lifetimes obtained for graphene flakes functionalized with Atto 465, Atto 425, and Atto 633 and for the respective pure fluorophores in chloroform (3 μg μL$^{-1}$). Average value ± standard deviation from five measurements.

| Sample | τ (ns) | $\tau_1$ (ns) | $\tau_2$ (ns) |
|---|---|---|---|
| Atto 465 (pure) | 3.5±0.1 | | |
| Atto 465-graphene flakes | | 0.8±0.1 | 2.9±0.1 |
| Atto 425 (pure) | 2.8±0.1 | | |
| Atto 425-graphene flakes | | 0.5±0.1 | 2.7±0.1 |
| Atto 633 (pure) | 3.7±0.1 | | |
| Atto 633-graphene flakes | | 0.5±0.1 | 2.6±0.1 |

**References:**

1  Ahmed Y.M.Z., El-Sheikh S.M., Zaki Z.I. Changes in hydroxyapatite powder properties via heat treatment. *Bull. Mater. Sci.* **2015**, *38,* 1807–1819. https://doi.org/10.1007/s12034-015-1047-0.

2  Rudolph W.W. Raman- and infrared-spectroscopic investigations of dilute aqueous phosphoric acid solutions, *Dalt. Trans.* **2010**, *39,* 9642–9653. https://doi.org/10.1039/C0DT00417K

3  Abifarin J.K., Obada D.O., Dauda E.T., Dodoo-Arhin D. Experimental data on the characterization of hydroxyapatite synthesized from biowastes. *Data Brief* **2019**, *26* 104485. https://doi.org/10.1016/j.dib.2019.104485.

4  Mecozzi M., Pietroletti M., Scarpiniti M., Acquistucci R., Conti M.E. Monitoring of marine mucilage formation in italian seas investigated by infrared spectroscopy and independent component analysis. *Environ. Monit. Assess.* **2012**, *184*, 6025–6036. https://doi.org/10.1007/s10661-011-2400-4.

5  Shao D., Wei Q., Microwave-assisted rapid preparation of nano-ZnO/Ag composite functionalized polyester nonwoven membrane for improving its UV shielding and antibacterial properties. *Materials* **2018**, 11. https://doi.org/10.3390/ma11081412.

6  Tretinnikov O.N., Zagorskaya S.A., Determination of the degree of crystallinity of poly(vinyl alcohol) by FTIR spectroscopy, *J. Appl. Spectrosc.* **2012**, *79*, 521–526. https://doi.org/10.1007/s10812-012-9634-y.

7   Bishop J., Murad E., Dyar M.D. The influence of octahedral and tetrahedral cation substitution on the structure of smectites and serpentines as observed through infrared spectroscopy. *Clay Minerals* **2002**, *37*, 617–628. https://doi.org/10.1180/0009855023740064.